\documentclass[prl,aps,twocolumn,groupedaddress,showpacs,floatfix]{revtex4}
\usepackage{amsmath,amssymb,multirow,epsfig,bm}

\newcommand{\etal}{{\it et al.,\;}}
\newcommand{\beq}{\begin{equation}}
\newcommand{\eeq}{\end{equation}}
\newcommand{\bea}{\begin{eqnarray}}
\newcommand{\eea}{\end{eqnarray}}

\newcommand{\veps}{\varepsilon}

\newcommand{\nn}{\nonumber}
\newcommand{\benn}{\begin{displaymath}}
\newcommand{\eenn}{\end{displaymath}}

\begin{document}

\title{ Local Density Functional Theory for Superfluid Fermionic Systems: The Unitary Gas }

\author{ Aurel Bulgac  }
\affiliation{Department of Physics, University of
Washington, Seattle, WA 98195--1560, USA}

\begin{abstract}
 
The first detailed comparison between {\it ab initio} calculations of finite fermionic superfluid systems, performed recently by Chang and Bertsch 
[Phys. Rev. A {\bf 76}, 021603(R), (2007)] and by von Stecher, Greene and Blume [e-print arXiv:00705:0671v1], and the extension of the density functional theory Superfluid Local Density Approximation (SLDA) is presented. It is shown that SLDA  reproduces the total energies, number density distributions in inhomogeneous systems along with the energy of the normal state in homogeneous systems. Unlike the Kohn-Sham LDA, in SLDA the effective fermion mass differs from the bare fermion mass and the spectrum of elementary excitations is also reproduced.
 
\end{abstract}

\date{\today}


\maketitle

The Density Functional Theory (DFT) introduced by Hohenberg and Kohn \cite{HK} became the tool of choice in the calculation of the properties of essentially most electron systems \cite{DG} after the introduction of the Local Density Approximation (LDA) by Kohn and Sham \cite{KS}. In order to achieve the accuracy needed in particular in chemical applications, a number of extensions of the LDA have been developed, the Local Spin Density Approximation (LSD(A)), the Generalized Gradient Approximation (GGA), etc., which have been thoroughly tested on a large variety of systems over the years by comparing the results of the LDA, LSD(A), and GGA with {\it ab initio} calculations and by refining the form of the density functionals used in practice \cite{DG,perdew}. All of these formulations rely on the Kohn-Sham orbitals and thus cannot deal effectively with superfluidity. The DFT extension to superfluid systems is a fundamental problem of the many-body theory.
Almost two decades ago such an extension was suggested 
\cite{gross}, however in terms of a non-local pairing field. A DFT formalism in terms of non-local fields is definitely intuitively less transparent, significantly harder to deal with in practice, and most likely physically not well motivated. The fact that the BCS theory leads formally to a non-local pairing field is not a strong argument in favor of such an approach. Such an argument was not used in the case of normal systems. All the BCS results could be recovered easily within a fully local formulation \cite{slda,g_eff}. There was a technical motivation to proceed with such a non-local formulation in the case of superfluid systems: the presence of a rather annoying divergence in the definition of a local anomalous density \cite{slda,g_eff}, an issue which has been successfully dealt with in Refs. \cite{slda,g_eff,nuclei}. This particular extension of the DFT to superfluid systems was dubbed Superfluid Local Density Approximation (SLDA) and it was applied so far to nuclear binding systematics and to  quantized vortices \cite{nuclei}. 

Here I shall analyze the properties of a fermion system at unitarity, when the strength of the interaction in a two-species fermion system (spin-up and spin-down) corresponds to an infinite scattering length \cite{GFB}. The basic properties of such a system, the energy per particle, pairing gap, etc. have been established in a series of {\it ab initio} calculations \cite{carlson,reddy,juillet,joe} for homogeneous systems. In Refs. \cite{CB,blume} two independent groups report on {\it ab initio} calculations of the properties of fermions at unitarity in a harmonic oscillator trap. These results provide the unique opportunity to test directly the accuracy of the SLDA, by deducing from the {\it ab initio} calculations of the homogeneous matter the appropriate local energy density functional and then use it to predict the properties of a fermion system in the presence of an external one-body potential. The proof of the Hohenberg and Kohn theorem is extended in a trivial manner \cite{DG} to the case when the external field has in second quantization the structure:
\beq
\sum_{\sigma=\uparrow,\downarrow} V_{ext}({\bf r})
\psi^\dagger_\sigma({\bf r})\psi _\sigma({\bf r})
+[ \Delta_{ext}({\bf r})\psi^\dagger_\uparrow({\bf r})\psi^\dagger_\downarrow({\bf r})
+ H.c. ].
\eeq
Thus one establishes that there is a unique mapping between the external potential, the total wave function of the system and the normal and anomalous densities and that a unique density functional of these densities exists. Consequently, the ground state energy of a superfluid system can be computed using a functional of the normal and anomalous densities. Here I shall analyze systems for which only 
$V_{ext}({\bf r})$ exists and one can consider formally that the external pairing field $\Delta_{ext}({\bf r})$ is vanishingly small, but not identically zero. This is assumed simply to force the reader to appreciate the fact that particle projection to a ``wave function with a well defined particle number" is neither required nor needed in order to recover the correct ground state energy.  

Fermions in the unitary regime (when the scattering length $a$ is large) are particularly attractive for a number of reasons: (1) this is a strongly interacting system which exhibits superfluid behavior and a complex phase diagram \cite{BF}; (2) at unitarity ($|a|=\infty$) the form of the energy density functional is restricted by dimensional arguments; (3) the availability  of {\it ab initio} results for homogeneous and inhomogeneous systems; (4) the relevance of this systems to a large variety of physical systems (low density neutron matter in neutron stars, cold fermionic atoms in traps, high-$T_c$ superconductivity, etc.); (5) tunability of the interaction both theoretically and experimentally. 

Dimensional arguments suggest at unitarity the simplest SLDA energy density functional (units $\hbar=m=1$):
\bea
&& {\cal E}({\bf r})= \alpha \frac{\tau({\bf r})}{2}
                 + \beta \frac{3(3\pi^2)^{2/3}n^{5/3} ({\bf r})}{10}
                 + \gamma \frac{|\nu({\bf r})|^2}{n^{1/3} ({\bf r})},\\
&&  n({\bf r})   = 2\sum_k |v_k({\bf r})|^2,\quad 
  \tau({\bf r}) = 2\sum_k |\nabla v_k({\bf r})|^2,\\
&& \nu({\bf r})  = \sum_k v_k^*({\bf r})u_k({\bf r}),
\eea
where $\alpha$, $\beta$ and $\gamma$ are dimensionless parameters and $n({\bf r})$,
$\tau({\bf r})$ and $\nu({\bf r})$ are the number/normal, kinetic and anomalous densities expressed through the usual Bogoliubov quasi-particle wave functions 
$[u_k({\bf r}), v_k({\bf r})]$ and where $k$ labels the quasi-particle states. The new universal parameter $\alpha$ is being introduced in this work and its presence proves to be critical in order to achieve the high accuracy demonstrated below. Since the kinetic and anomalous densities diverge \cite{slda,g_eff} one has to introduce a renormalization procedure for the pairing gap and for the energy density. The renormalized density functional and the equations for the quasi-particle wave functions obtained by the standard variation are as follows:
\bea
&& \!\!\!\!\!\!\!\!\!\!\!\!\!\!
{\cal E}({\bf r}) =\alpha \frac{\tau_c({\bf r})}{2}
                 + \beta \frac{3(3\pi^2)^{2/3} n^{5/3}({\bf r})}{10} 
                 + g_{\it eff}({\bf r})|\nu_c({\bf r})|^2   \nn\\
&& \!\!\!\!\!\!\!\!\!
 +V_{ext}({\bf r})n({\bf r}), \label{eq:energy}\\
& & \frac{1}{ g_{{\it eff}}({\bf r})}=
\frac{n^{1/3}({\bf r})}{\gamma} +\Lambda_c({\bf r}), \label{eq:geff} \\
&& \!\!\!\!\!\!\!\!\!\!\!\!\!\!
\tau_c({\bf r}) = 2\!\!\!\sum_{E_k<E_c} |\nabla v_k({\bf r})|^2, \quad
\nu_c({\bf r})  = \sum_{E_k<E_c} v_k^*({\bf r})u_k({\bf r}),\\
&& \!\!\!\!\!\!\!\!\!\!\!\!\!\!
\left \{ \begin{array}{l}
[h({\bf r})-\mu]u_k ({\bf r})+\Delta({\bf r})v_k ({\bf r})= E_k u_k ({\bf r}), \\ 
 \Delta^*({\bf r})u_k({\bf r})-[ h({\bf r})-\mu ]v_k({\bf r})=E_k v_k ({\bf r}),
\end{array}
\right . \label{eq:bdg}\\
&& \!\!\!\!\!\!\!\!\!\!\!\!\!\!
h ({\bf r})= -\frac{\alpha \nabla^2}{2} 
+ U({\bf r}),   \\
& & \!\!\!\!\!\!\!\!\!\!\!\!\!\!
U({\bf r})= 
\frac{\beta(3\pi^2n({\bf r}))^{2/3}}{2} 
   -\frac{|\Delta({\bf r})|^2}{3\gamma n^{2/3}({\bf r})}+V_{ext}({\bf r}),
\label{eq:pot} \\
& & \!\!\!\!\!\!\!\!\!\!\!\!\!\!
\Delta({\bf r}):= -g_{{\it eff}}({\bf r})\nu_c({\bf r}), \\
&&  \!\!\!\!\!\!\!\!\!\!\!\!\!\!
\Lambda_c({\bf r})=-\frac{ k_c({\bf r}) }{ 2\pi^2\alpha }
 \left \{ 1 -\frac{ k_0({\bf r}) }{ 2 k_c({\bf r}) }
\ln \frac{ k_c({\bf r})+k_0({\bf r}) }{ k_c({\bf r})-k_0({\bf r}) }
    \right \}    \label{eq:Lambda} \\
& &  \!\!\!\!\!\!\!\!\!\!\!\!\!\! 
E_c+\mu = \frac{\alpha k_c^2({\bf r})}{2} + U({\bf r}), \quad
     \mu = \frac{\alpha k_0^2({\bf r})}{2} + U({\bf r}). 
\eea
Here $E_c$ is an energy cutoff and I have added as well an external potential 
$V_{ext}({\bf r})$. It can be shown that in the total energy the kinetic and anomalous densities have to enter in the combination
\beq
\alpha \frac{\tau_c({\bf r})}{2}-\Delta({\bf r})\nu_c({\bf r})\equiv 
\alpha \frac{\tau_c({\bf r})}{2}+ g_{\it eff}({\bf r})|\nu_c({\bf r})|^2
\eeq
and thus the pairing part of the functional is uniquely defined. The second term in Eq. (\ref{eq:pot}) for $U({\bf r})=\delta {\cal E}({\bf r})/\delta n({\bf r})$ is obtained by varying $g_{{\it eff}}({\bf r})$, see  Eq. (\ref{eq:geff}), with respect to 
$n({\bf r})$ and neglecting in the first approximation the dependence of $\Lambda_c({\bf r})$ on $U({\bf r})$. There is still a small correction term to $U({\bf r})$, arising from varying $\Lambda_c({\bf r})$, which can be made rather small if $E_c$ is sufficiently large. This additional term can be computed using
\beq
\frac{\delta \Lambda_c({\bf r})}{\delta n({\bf r})}=
-\frac{1}{4\pi^2\alpha^2k_0({\bf r})}
\ln \frac{k_c({\bf r})+k_0({\bf r})}{k_c({\bf r})-k_0({\bf r}) } 
\frac{\delta U({\bf r})}{\delta n({\bf r})}
\eeq
and one can argue that to a good approximation $\delta U({\bf r})/\delta n({\bf r})\approx 2(U({\bf r})-V_{ext}({\bf r}))/3n({\bf r})$, which is consistent with the universality of a homogeneous unitary Fermi gas (UFG), where $U({\bf r})\propto n^{2/3}({\bf r})$. The results are independent of the value of the cutoff $E_c$ if this is chosen appropriately large \cite{appendix}. The above formulas apply to systems with an even particle number. In order to describe systems with an odd particle number one has to place an extra quasiparticle in a specific quantum state $n_0$ \cite{appendix}. 
 
\begin{figure}[tbh]
\epsfxsize=9.0cm
\centerline{\epsffile{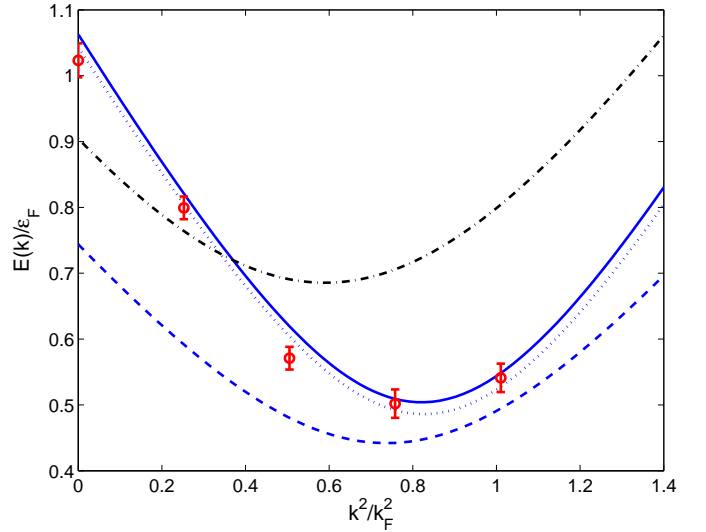}}
\caption{ \label{fig:spectrum} (Color online) The {\it ab initio} fermionic quasiparticle spectrum determined in Ref. \cite{reddy} ( red circles) compared with the SLDA spectrum defined by the {\it ab initio} parameters $\xi,\eta$ from Ref. \cite{reddy} (solid blue line), \cite{carlson} (dotted blue line), \cite{juillet} (dashed blue line) and naive BCS approximation at unitarity (black dotted-dashed line).}
\end{figure}

By requiring that a homogeneous gas of number density $n=N/V=k_F^3/3\pi^2$ has an energy per particle $E/N = 3 \xi_S \veps_F/5$, a chemical potential $\mu = \xi_S\veps_F$ and a pairing gap $\Delta =\eta \veps_F$, where $\veps_F=k_F^2/2$, one can determine the dimensionless parameters $\alpha$, $\beta$ and $\gamma$ in Eq. (\ref{eq:energy}).  The corresponding equations determining these parameters are:
\bea
&&n=\frac{k_F^3}{3\pi^2}
 =\int \frac{d^3k}{(2\pi)^3}\left ( 1-\frac{\veps_k}{E_k}\right ),\label{eq:alpha}\\
&& \frac{3}{5}\veps_F n\xi_S =  \nn \\
&& \int \frac{d^3k}{(2\pi)^3}
\left [ \frac{\alpha k^2}{2}\left ( 1 -\frac{\veps_k}{E_k}\right )
-\frac{\Delta^2}{2E_k}\right ] + \frac{3}{5}\veps_F n\beta,  \label{eq:beta}\\
&& \frac{n^{1/3}}{\gamma}=\int \frac{d^3k}{(2\pi)^3}
\left ( \frac{1}{\alpha k^2}-\frac{1}{2E_k} \right ), \label{eq:gamma}
\eea
where
\bea
&&\veps_k =\frac{\alpha k^2}{2}+U-\mu=
  \frac{\alpha k^2}{2}+({\overline \beta}-\xi_S)\veps_F, \\
&& E_k=\left ( \veps_k^2+\Delta^2\right ) ^{1/2},  \quad
{\overline \beta} =\beta-\frac{(3\pi^2)^{2/3}\eta^2}{6\gamma},
\eea
The momentum $k_F$ can be factored out of the equations [\ref{eq:alpha}, \ref{eq:beta}), and (\ref{eq:gamma}], as expected for a system at unitarity.
Using $\xi_S = 0.42(2)$ and $\eta = 0.504(24)$ determined by Carlson and Reddy \cite{reddy} one obtains $\alpha= 1.14$, $\beta = -0.553$ and $1/\gamma=-0.0906$. 
Using the original values of Carlson \etal \cite{carlson}, namely $\xi_S=0.44$ and 
$\eta=0.486$ one obtains instead $\alpha=1.12$, $\beta=-0.520$ and 
$1/\gamma=-0.0955$. A different set of values has been calculated recently by Juillet \cite{juillet}, $\xi_S=0.449(9)$ and $\eta=0.442(3)$, which lead to 
$\alpha = 0.812$, $\beta = -0.172$ and $1/\gamma =-0.0705$.  
The value of $\alpha=1.14$ leads to an effective mass of $m_{\it eff}/m=1/\alpha = 0.877$. 

One can now calculate the spectrum of the elementary fermionic excitations of a homogeneous UFG, shown in Fig. \ref{fig:spectrum}, and compare it with the {\it ab initio} spectrum determined in Ref. \cite{reddy}. The agreement between the two spectra is nothing short of spectacular, especially if one has in mind that DFT is not usually expected to reproduce the single-particle spectrum \cite{HK,DG,KS}, unless one sets up from the outset to achieve that as well \cite{furnstahl,blm}. The value of the effective mass is a matter of mild surprise, and also the fact that the minimum of $E_k$ occurs at $k_0=0.906k_F$. The analysis of Ref. \cite{son} arrives at a similar value for $k_0$. 
The energy per particle of a homogeneous UFG in the normal state has also been determined \cite{carlson,joe}, $E_N/N = 3 \xi_N \veps_F/5$ with $\xi_N=0.55$, which is an average of the two results. The SLDA energy density functional is consistent with this value, as $\xi_N=\alpha+\beta = 0.59$.  The parameters $\xi$ and $\eta$ from Ref. \cite{juillet}  lead to $\xi_N=0.64$ and an effective mass $m_{\it eff}/m=1/\alpha = 1.23$ and to a quasi-particle spectrum in strong disagreement with the results of Ref. \cite{reddy}.

\begin{figure}[tbh]
\epsfxsize=8.0cm
\centerline{\epsffile{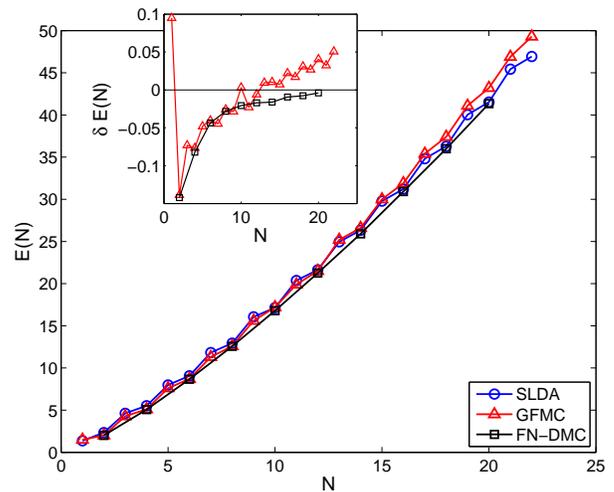}}
\caption{ \label{fig:energy}  (Color online) The comparison between the GFMC \cite{CB}, FN-DMC 
\cite{blume} and SLDA total energies $E(N)$.  The clear odd-even staggering of the energies is due to the onset of the pairing correlations. The inset shows the discrepancy between the GFMC and FN-DMC and SLDA energies, $\delta E(N)=E_{MC}(N)/E_{SLDA}(N)-1$, where $E_{MC}(N)$ stands for the energies obtained in GFMC or FN-DMC respectively. }
\end{figure}

\begin{figure}[tbh]
\epsfxsize=9.0cm
\centerline{\epsffile{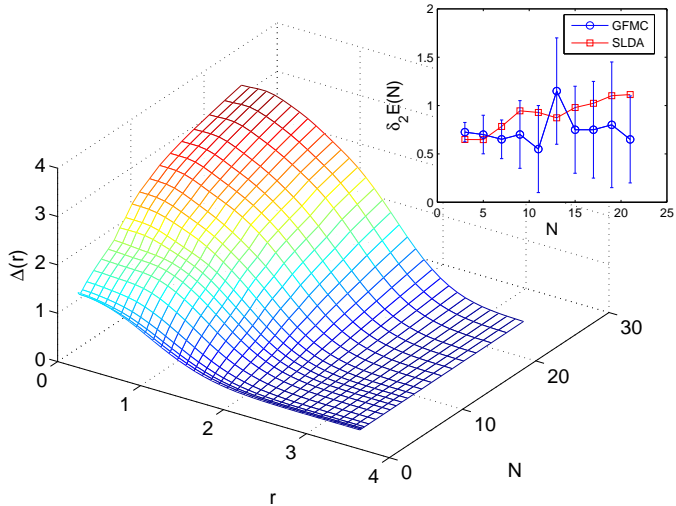}}
\caption{ \label{fig:pairing} (Color online)  The pairing field $\Delta({\bf r})$ for even particle number systems. The inset shows the quantity $\delta_2 E(N)=E(N)-(E(N+1)+E(N-1))/2$ calculated in SLDA (red squares) and GFMC (blue circles).  }
\end{figure}

The SLDA calculations for the finite systems [with $V_{ext}({\bf r}) = m\omega^2r^2/2$] have been performed using the Bessel Discrete Variable Representation (DVR) method \cite{dvr}, and a few details are presented in Ref. \cite{appendix}. The comparison between the Green Function Monte Carlo 
(GFMC) \cite{CB}, Fixed Node-Diffucion Monte Carlo 
(FN-DMC) \cite{blume} and SLDA is presented in Figs. \ref{fig:energy},\ref{fig:pairing} and in Ref. \cite{appendix} (units for energy and length defined by $\hbar=m=\omega=1$). 
The agreement between the Monte-Carlo (MC) and SLDA results is very good, especially keeping in mind that the MC calculations for both infinite matter and finite systems have aside from statistical errors also noticeable systematic errors.  Both GFMC \cite{CB} and FN-DMC \cite{blume} calculations are in principle variational and, since the energies for the larger systems in the FN-DMC calculation are consistently lower that the corresponding GFMC results, one can expect that the FN-DMC results are somewhat more accurate. That is also in line with the smooth behavior of the $\delta E(N)= E_{FN-DMC}(N)/E_{SLDA}(N)-1$ with $N$ as opposed to the behavior of $\delta E(N)= E_{GFMC}(N)/E_{SLDA}(N)-1$. The SLDA energies converge unexpectedly fast towards the FN-DMC values, 
see Fig. \ref{fig:energy} and Table I \cite{appendix}. As a reference, the Thomas-Fermi energy $E(N)=(3N)^{4/3}\sqrt{\xi_S}/4$, which for $N=20$ comes to 38.05 as compared to 43.2, 41.35 and 41.51 in GFMC, FN-DMC, and SLDA respectively. This discrepancy, whose size is typical in this particle range, is comparable in magnitude with the condensation energy, but opposite in sign.  

As far as the density profiles are concerned, they agree reasonably well in the surface region, but show noticeable differences in the central region 
\cite{appendix}. The reasons for this disagreement are likely the relatively poorer quality of the GFMC results for larger particle numbers \cite{blume}. It is notable that there is a small discrepancy between the SLDA energies calculated as expectation values of the functional and using the viral theorem \cite{virial}, a discrepancy which seems to decrease with increasing $E_c$. Notice that there is a somewhat bigger discrepancy in the GFMC values of the energies and the corresponding virial expectations \cite{CB,appendix}.

Even though the agreement between the SLDA and {\it ab initio} results is surprisingly good, a better agreement would have been bad news. The reasons for the discrepancies can be ascribed to several origins: 
{\it (i)} The energy density functional assumed here, see Eq. (\ref{eq:energy}), is certainly not unique and further efforts should be devoted to study other possible forms. {\it (ii)} The so called Self-Interaction Correction (SIC) \cite{DG} is not present here, and its absence is seen in the SLDA energy for $N=1$, which is $1.37\hbar\omega$ instead of $1.5\hbar\omega$. Even though $\delta_2 E(N)=E(N)-[E(N+1)+E(N-1)]/2$ calculated in the two methods agree within the error bars, some differences are likely due to {\it (iii)} the absence of the polarization of the even core by the field of extra odd particle (full spherical symmetry was assumed). Another reason is {\it (iv)} the absence of spin number densities in this formulation of the SLDA and an extension to Superfluid LSD approximation is needed, especially for small systems. For now, the energies of the homogeneous asymmetric systems are known with significantly less accuracy, see Ref. \cite{BF}. And last, but not least, {\it (v)} gradient correction terms are needed (beyond those already included through the use of explicit single-particle kinetic energy). 

The differences between the {\it ab initio} and the SLDA results are noticeably less than the corresponding differences in electronic systems and the validation of the DFT extension presented here is to my knowledge a first of such kind.   
I anticipate that the existence of an accurate SLDA will have a great impact on several physics subfields: atomic nuclei, neutron matter and quark matter in neutron stars, dilute atomic Fermi gases, condensed matter and, maybe, on high $T_c$ superconductivity as well.  The ability to perform high accuracy calculations using essentially meanfield techniques in lieu of {\it ab initio} calculations can hardly be underestimated.  


I am grateful to S.Y. Chang and G.F. Bertsch for the use of their unpublished results 
and for discussions, J. Carlson for the data from Ref. \cite{reddy}, to  M.M. Forbes for numerous discussions and help with coding and numerics. A number of discussions and the valuable comments of W. Kohn are particularly appreciated and I thank D. Blume for information about their unpublished results. Support is acknowledged from the DOE under grants DE-FG02-97ER41014 and DE-FC02-07ER41457.

{\em Notes added in proof.}  The recent FN-DMC calculation of D. Blume, J. von Stecher, and C.H. Greene, arXiv:0708:2734v1, extended to both even and odd systems up to $N=30$, demonstrated very nice agreement with SLDA. The Galilean invariance of Eqs. (2) and (5) becomes manifest 
upon adding the term $(1-\alpha){\bm p}^2({\bm r})/2n({\bm r})$, where 
${\bm p}({\bf r})= \text{Real} [-i\sum_kv^*_k({\bm r}){\bm \nabla} v_k({\bm r})]$.   
 

\newpage

\vspace{1cm}
{\bf Online appendix}

\vspace{1cm}

The emerging SLDA equations Eqs. (\ref{eq:bdg}) have been discretized using the Bessel Discrete Variable Representation (DVR) method outlined in Ref. \cite{dvr}. I have used two DVR bases, one with angular momentum $l=0$ for all even angular momenta and with $l=1$ for odd angular momenta. In this representation only a part of the kinetic energy is a non-diagonal matrix, while both the meanfield and the pairing fields are diagonal. For angular momenta $l>1$ either $l(l+1)/2r^2$ for even $l$'s and $[l(l+1)-2]/2r^2$ for odd $l$'s was included into the potential. The number of discrete points was  up to $N_{max}=100$, which is equivalent to the use of an harmonic oscillator basis with at least an equal number of oscillator shells. Larger basis sets have been used as well. Comparable results were obtained with a basis set of 35 points.  However, unlike the decomposition of the quasi-particle wave functions in a harmonic oscillator basis, which has significant convergence problems, the use of Bessel DVR method is noticeable faster converging and for many quantities one can relatively easily reach relative machine precision ($\approx 1.0e-15$). The iterative process was also significantly improved by the use of the Broyden method in determining the meanfield, pairing field and chemical potentials. A fully converged solution, at the level of 1.0e-9, was obtained in less than 20-30 iterations, irrespective of particle number. The accuracy is typically improved by an order of magnitude every 2-3 iterations, once the process starts converging. All calculations were performed on a laptop, using a MATLAB program of about 400 lines, including the plotting and the comparison with GFMC/FN-DMC calculations. The part which solves the SLDA equations alone is about 200 lines. For a basis set of $N_{max}=35$ points, a fully self-consistent solution for a given particle number is obtained in about 6 seconds, while for a basis set of $N_{max}=75$ points the time increases to about 30 seconds and 55 seconds on average for $N_{max}=100$. Odd systems usually require more iterations. The program spends most of the time (more than 90 \%)  diagonalizing the SLDA equations (\ref{eq:bdg}).  A number of details can be found at {\tt $www.phys.washington.edu/users/bulgac/$ $Media/MMFAB.pdf$}.

In order to describe systems with an odd particle number one has to place a quasi-particle in a specific quantum state $n_0$ and change the densities according to the following prescription:
\bea
&& \!\!\!\!\!\!\!\!\! n({\bf r})      = 2\sum_n |v_n({\bf r})|^2 +  
   |u_{n_0}({\bf r})|^2 - |v_{n_0}({\bf r})|^2, \\
&& \!\!\!\!\!\!\!\!\! \tau_c({\bf r}) = 2\sum_n |\nabla v_n({\bf r})|^2+  
   |\nabla u_{n_0}({\bf r})|^2 - |\nabla v_{n_0}({\bf r})|^2,\\
&& \!\!\!\!\!\!\!\!\! \nu_c({\bf r})  = \sum_n v_n^*({\bf r})u_n({\bf r}) - 
    v_{n_0}({\bf r})u_{n_0}^*({\bf r}).
\eea
It can be shown that this corresponds to exactly one particle difference between the spin-up and spin-down particle numbers. It also amounts to the fact that the single-particle quantum state $n_0$ is excluded from the pairing mechanism. This is often referred to as blocking and it is the reason for the staggering in the ground state energies of the even and odd systems. Essentially identical results can be obtained by introducing different chemical potentials for the spin-up and spin-down components.

\begin{table}[tbh]
\caption{Table I. The energies $E(N)$ calculated within the GFMC \cite{CB}, FN-DMC \cite{blume} and SLDA. When two numbers are present the first was calculated as the expectation value of the Hamiltonian/functional, while the second is the value obtained using the virial theorem, namely $E(N)= m\omega^2\int d^3rn({\bf r})r^2$  \cite{virial}.} 
\begin{tabular}{||l|l|l|l||} \hline \hline
$N$    & $E_{GFMC}$ & $E_{FN-DMC}$&  $E_{SLDA}$\\
\hline \hline
 1 &   1.5          &      & 1.37        \\ \hline
 2 &   2.01/1.95    &2.002 & 2.33/2.34   \\ \hline 
 3 &   4.28/4.19    &      & 4.62/4.62   \\ \hline
 4 &   5.10         &5.069 & 5.52/5.56   \\ \hline
 5 &   7.60         &      & 7.98/8.02   \\ \hline
 6 &   8.70         &8.67  & 9.07/9.14   \\ \hline 
 7 &   11.3         &      & 11.83/11.91 \\ \hline 
 8 &   12.6/11.9    &12.57 & 12.94/13.06 \\ \hline
 9 &   15.6         &      & 16.06/16.20 \\ \hline
10 &   17.2         &16.79 & 17.15/17.33 \\ \hline
11 &   19.9         &      & 20.36/20.56 \\ \hline
12 &   21.5         &21.26 & 21.63/21.88 \\ \hline
13 &   25.2         &      & 24.96/25.23 \\ \hline 
14 &   26.6/26.0    &25.90 & 26.32/26.65 \\ \hline
15 &   30.0         &      & 29.78/30.14 \\ \hline
16 &   31.9         &30.92 & 31.21/31.62 \\ \hline
17 &   35.4         &      & 34.81/35.26 \\ \hline 
18 &   37.4         &36.00 & 36.27/36.78 \\ \hline 
19 &   41.1         &      & 40.02/40.58 \\ \hline
20 &   43.2/40.8    &41.35 & 41.51/42.12 \\ \hline
21 &   46.9         &      & 45.42/46.10 \\ \hline
22 &   49.3         &      & 46.92/47.64 \\ \hline \hline

\end{tabular}
\end{table}

The SLDA energies and the corresponding GFMC and FN-DMC results are presented in Table I for a DVR basis set using $N_{max}=100$ points. In a calculation using $N_{max}=35$ points one observes differences typically at the level of 1.0e-4 or less for the energies, which can be ascribed to the weak dependence of the results on the cutoff energy $E_c$. However, the agreement between the energy expectation values and the values calculated using the virial theorem degrades as $N_{max}$ is decreased. M.M. Forbes noticed that using a slightly smoothed cutoff procedure eliminates a significant part of the energy cutoff dependence. For odd systems in particular we found that performing the calculations at a very small temperature $T = 0.01\cdots 0.1\hbar\omega$ eliminates the need to guess the optimal quantum numbers $n_0$ of the extra quasi-particle.

\begin{figure}[tbh]
\epsfxsize=5.5cm
\centerline{\epsffile{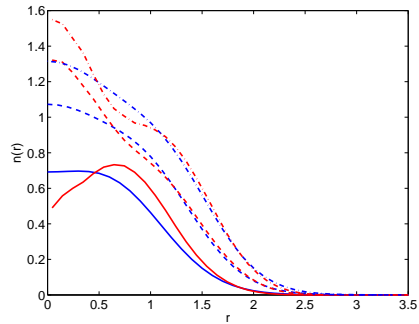}}
\caption{ \label{fig:density}  (Color online)
The number densities for $N=8$ (solid line), 14 (dashed line) and 20 (dot-dashed line), calculated within SLDA (blue line) and GFMC (red line), units as in Fig. \ref{fig:energy}. There are no shell effects in the SLDA number densities and their presence in the GFMC results is probably due partially to statistical reasons and/or partially to a less than optimal structure of the trial wave function. }
\end{figure}

The number densities show a reasonable agreement in the surface region. However, the GFMC number densities, unlike the SLDA number densities, show pronounced shell effects in the interior region. The magnitude of the pairing field for the smaller systems is such that it would not allow for an efficient pairing across harmonic oscillator shells.  The odd-even staggering of the total energies is however equally well manifest over the entire range of particle number.

\end{document}